%% file: main.tex
\documentclass[14pt,twocolumn]{aastex631}
\usepackage{threeparttable}
\usepackage{color}
\usepackage{multirow}

\usepackage{amsmath}

\graphicspath{{./}{figures/}}
\usepackage{subfiles}
\usepackage{subfigure}
\submitjournal{ApJ Letters}

\shorttitle{Search for Neutrino Emission from X-ray Binaries}
\shortauthors{IceCube Collaboration}

\begin{document}

\email{analysis@icecube.wisc.edu}
\title{Search for High-Energy Neutrino Emission from Galactic X-ray Binaries with IceCube}

\input{authors.tex}

\begin{abstract}
We present the first comprehensive search for high-energy neutrino emission from high- and low-mass X-ray binaries conducted by IceCube. Galactic X-ray binaries are long-standing candidates for the source of Galactic hadronic cosmic rays and neutrinos. The compact object in these systems can be the site of cosmic-ray acceleration, and neutrinos can be produced by interactions of cosmic rays with radiation or gas, in the jet of a microquasar, in the stellar wind, or in the atmosphere of the companion star. We study X-ray binaries using 7.5 years of IceCube data with three separate analyses. In the first, we search for periodic neutrino emission from 55 binaries in the Northern Sky with known orbital periods. In the second, the X-ray light curves of 102 binaries across the entire sky are used as templates to search for time-dependent neutrino emission. Finally, we search for time-integrated emission of neutrinos for a list of 4 notable binaries identified as microquasars. In the absence of a significant excess, we place upper limits on the neutrino flux for each hypothesis and compare our results with theoretical predictions for several binaries. In addition, we evaluate the sensitivity of the next generation neutrino telescope at the South Pole, IceCube-Gen2, and demonstrate its power to identify potential neutrino emission from these binary sources in the Galaxy.
\end{abstract}


\section{Introduction} \label{sec:intro}
Cosmic rays (CRs) up to several PeV, the ``knee'' in the CR spectrum, are believed to be of Galactic origin. However, where and how these CRs are accelerated remains an open question. Interactions of hadronic CRs in the Galaxy will lead to the production of pions, which subsequently decay into gamma rays and neutrinos with energies potentially reaching hundreds of TeV. Unlike gamma rays which could also be produced by accelerated electrons/positrons, high-energy neutrinos would be the smoking-gun for CR interactions as they are the only way to produce neutrinos. In 2013, IceCube discovered TeV-PeV neutrinos of astrophysical origin~\citep{Aartsen:2013jdh,Aartsen:2014gkd}. Nevertheless, the sources of those high-energy neutrinos are yet to be identified. Since the discovery of high-energy cosmic neutrinos by IceCube, many studies have been conducted searching for neutrino emission from point-like sources, extended regions, and the diffuse emission from CRs interacting with the Galactic interstellar medium. The isotropic distribution of neutrino events in IceCube suggests a dominant contribution from extragalactic sources and constrains the Galactic contribution to the diffuse neutrino flux to less than 14\% above 1 TeV~\citep{Aartsen:2017ujz}.

Recent searches for correlations with Galactic sources show no evidence of a signal, e.g., in \cite{Albert:2018vxw,Kheirandish:2019bke,Aartsen:2020eof} and so far, only a blazar, TXS 0506+056, which is extragalactic, shows evidence of being a neutrino source~\citep{IceCube:2018cha,IceCube:2018dnn}. 

X-ray binaries (XRBs) are binary systems that emit X-rays and consist of a compact object (neutron star (NS) or black hole (BH)) and a non-compact companion star. The mass from the companion is accreted onto the compact object due to the strong gravitational attraction forming an accretion disk. Depending on the mass of the companion star, XRBs can be divided into high-mass XRBs (HMXBs) and low-mass XRBs (LMXBs). The companion star of an HMXB is massive, usually an O or B-type star, and the accretion is mainly a result of stellar wind capture~\citep{Liu:2007tu}. LMXBs have companion stars of A-type or later, and the mass transfer is mostly caused by Roche lobe overflows~\citep{Liu:2007ts}. These systems are bright in X-rays and sometimes in gamma rays. XRBs have been proposed as sites of CR acceleration and hadronic interactions since the 1980s~\citep[e.g.,][]{Gaisser:1983cj,Berezinsky:1985xp}. Microquasars, which are XRBs with relativistic jets, are miniature analogs of quasars, a category of extremely luminous active galactic nuclei. It is expected that the similarity may not only be morphological, but also in the physics of the formation of the accretion disk and the behavior of the jet, at different luminosity scales. Hadronic processes in jets of microquasars have been widely discussed. Protons can be accelerated in the jet, and pions are generated through interactions with the external radiation field of the accretion disk and/or internal synchrotron photons ($p\gamma$)~\citep{Levinson:2001as,Distefano:2002qw,Romero:2008qm}. Other scenarios focus on hadronuclear interactions ($pp$), e.g., jet-cloud/wind interactions when the jet traversing the matter field of the ejected clouds or stellar wind from the companion star~\citep{aharonian1991model, Romero:2003td,Bednarek:2005gf}. For other XRBs without collimated beams, hadronic interactions happen in a wider shocked region. CR can be accelerated in the magnetosphere of a spinning NS and then interact with matter from either the accretion disk or the companion star~\citep{Gaisser:1983cj,Kolb:1985bb,Berezinsky:1985xp,cheng1989period,Anchordoqui:2002xu}. 

Some XRBs have been observed in TeV gamma rays, which may suggest acceleration of hadrons besides an origin of accelerated leptons. Neutrino production in those sources and the potential of IceCube detection have been discussed, e.g., LS 5039~\citep{Aharonian:2005cx}, LS I +61 303~\citep{Christiansen:2005gw,Torres:2006ub}, and SS 433~\citep{Reynoso:2019vrp,Kimura:2020acy}.

XRBs exhibit both outbursts and periodic emissions. It is reasonable to hypothesize that the possible neutrino emission is related to either the periodicity or the X-ray outbursts. Conducting time-dependent studies incorporating these hypotheses provides the advantage of suppressing background. Both time-integrated and time-dependent analyses searching for neutrino emission from the entire sky have been performed by IceCube~\citep{Abbasi:2011ke,Aartsen:2013uuv,Aartsen:2014gkd,Aartsen:2015wto} and ANTARES~\citep{Adrian-Martinez:2014ito,Albert:2016gtl} previously, without significant signal detection. Here, we present a comprehensive high-energy neutrino search focusing on XRBs using 7.5 years of IceCube data. This study explores the periodic, X-ray outburst-correlated, and persistent emission scenarios of the possible neutrino flux from XRBs, while covering a broader list of sources compared to previous studies.

\section{Detector \& Data Set}\label{sec:data}
The IceCube Neutrino Observatory, built under the ice surface at the South Pole, is a 1 $\rm{km}^3$ Cherenkov neutrino telescope. The detector is composed of an array of digital optical modules (DOMs), each equipped with a photomultiplier tube (PMT) and on-board read-out electronics. 
The PMTs in the DOMs collect the Cherenkov photons emitted by the relativistic charged particles produced in neutrino interactions when traversing the ice. The number of detected photons and their arrival time information are used to reconstruct the energy and direction of each event. Different neutrino flavors and interactions can result in different event signatures inside IceCube detection volume. Among them, muon tracks from $\nu_\mu/\bar{\nu}_\mu$ charged-current interactions can be reconstructed with a good angular resolution, typically $\lesssim 1^\circ$ \citep{IceCube:2016zyt}, which makes them ideal for neutrino point source searches. 

In this search, we use 7.5 years of all-sky muon track data collected between 2011-05-13 and 2018-10-14, with an effective livetime of 2711 days. The data sample being used has an event selection focusing on high-quality through-going neutrino track events from the entire sky with an energy threshold near 100~GeV. The final event rate is $\sim$6 mHz, yielding a total of 1,502,612 events. A better sensitivity in the Northern Hemisphere is expected due to the suppression of the atmospheric muon background by Earth for up-going events. Details of the data sample are described in \cite{Aartsen:2016lmt}. 

\begin{deluxetable*}{llDDDD}[t!]
\tablecaption{Table of the most significant sources}
\tablehead{
\colhead{Analysis} &\colhead{Source} & \multicolumn2c{TS} & \multicolumn2c{$\hat{n}_{s}$} & \multicolumn2c{$\hat{\gamma}$} & \multicolumn2c{$p$-value}
}
\decimals
\startdata
Periodic        & V635 Cas & 9.07 & 50.5  & 4        & 0.25$\;$(0.0052)       \\ 
Flare      & V404 Cyg & 8.28 & 5.4  & 4        & 0.75$\;$(0.014)         \\
Time-integrated & Cyg X-3  & 6.81 & 44.6  & 3.25     & 0.036$\;$(0.009)          \\
\enddata
\tablecomments{The most significant source in each analysis with its best-fitted TS, $\hat{n}_s$, $\hat{\gamma}$. Both post-trial and pre-trial (bracketed) $p$-values are shown. The trial correction is calculated by considering the number of sources studied in each analysis respectively and has a relation $p_{\mathrm{post}}=1-(1-p_{\mathrm{pre}})^N$, where N is the source number.} 
\label{tab:sig_results}
\end{deluxetable*}

\vspace{-1cm}
\section{Analysis} \label{sec:analysis}
The three analyses use an unbinned maximum likelihood-ratio method \citep{Braun:2008bg, Braun:2009wp} to search for an excess of neutrino events above the background of atmospheric muon, atmospheric and isotropic astrophysical neutrino flux in the direction of selected XRBs. In all analyses, the likelihood function describing the signal includes both a directional correlation and an energy spectrum assumed to follow a power law, i.e., $\propto E^{-\gamma}$. For the time-dependent analyses, each incorporates a unique temporal term in the likelihood to model the signal. As the data is expected to be background-dominated, which is uniform in time, the background probability density function (PDF) is constructed from time-randomized data. The test statistic (TS) is obtained by maximizing the likelihood ratio with respect to a set of parameters, which include the number of signal events ($n_s$) and a power-law spectral index ($\gamma$). Other parameters are analysis-specific, e.g., time of the neutrino emission and the duration. A more detailed description of the method can be found in the appendix Sec.~\ref{sec:method_details}. The sources studied are from a catalog of 114 Galactic HMXBs~\citep{Liu:2007tu} and 187 LMXBs~\citep{Liu:2007ts}. Furthermore, we added 7 TeV XRBs from TeVCat\footnote{\url{http://tevcat.uchicago.edu}} very-high-energy gamma-ray catalog~\citep{wakely2008tevcat}, which are not in the HMXB or LMXB catalog. Additional selection criteria are then applied to this initial source list in the two time-dependent analyses (as explained in Sec.~\ref{subsec:method_1} and Sec.~\ref{subsec:method_2}), leading to two different source lists with an overlap. We also include a time-integrated search on four individual sources. The rest of this section includes the description of each analysis and its results. Table~\ref{tab:sig_results} summarizes the most statistically significant source in each analysis. 

\begin{figure*}[t!]
    \centering
    \subfigure{\includegraphics[width=0.95\textwidth]{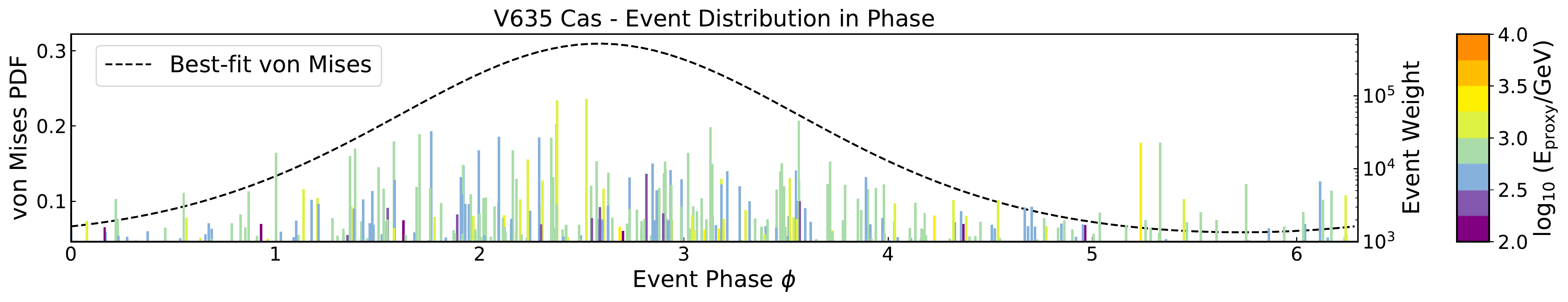}} %
    \vspace{0em}
	\subfigure{\includegraphics[width=0.95\textwidth]{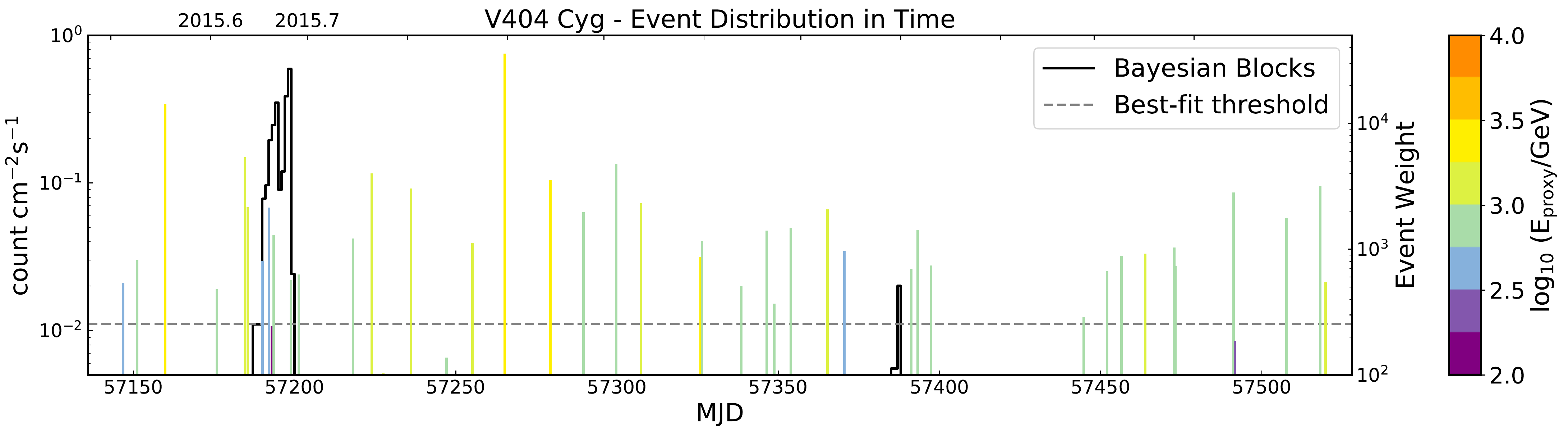}}
    \caption{{\bf{\textit Top:}} The time PDF and the event distribution within a $10^\circ$ zenith band centered at V635 Cas covering the whole 7.5 yr dataset with the time converted to the phase $\phi$ according to the best-fit parameters. Vertical lines represent neutrino events. The color shows the energy proxy while the height shows the weight of each event in the likelihood function, assuming an $E^{-2}$ spectrum. {
    \bf{\textit Bottom:}} The time PDF before normalization (Bayesian blocks) and the event distribution within $1.5^\circ$ around V404 Cyg in the data sample of 2015 at the time indicated by MJD. The major X-ray flare happened in the June of 2015. The Bayesian blocks have been shifted by the best-fit time lag and the dashed gray line indicates the best-fit threshold. The vertical lines are read the same way as in the top panel.}
    \label{fig:event_dist}
\end{figure*}

\subsection{Periodic Analysis}\label{subsec:method_1}
Periodic searches have been previously performed by IceCube using data of 2007-2009 with a partially-built configuration and 2008-2012, including the first year of the completed detector ~\citep{Abbasi:2011ke,Aartsen:2015wto}, assuming that the neutrino emission from XRBs is modulated by orbital periods. In the period-folded phase space, these neutrinos should appear as at the same phase. As an improvement on prior work, the temporal signal PDF in the phase space is modeled by a von Mises distribution instead of a Gaussian distribution (see Sec.~\ref{subsec:periodic_method_more} for details). The choice of the von Mises distribution is to satisfy the periodic boundary condition imposed by the wrapped event phase. Modeling the neutrino emission with an aperiodic Gaussian can cause a loss of statistical power, especially if the emission phase profile is wide and/or if the peak is close to the boundary. The parameters to be fitted in the temporal term are the phase peak, $\Phi_0$, the concentration, $\kappa$ which represents the spread of the events in phase, and the orbital period, $P$. A Gaussian prior centered at the measured period, $P_\mathrm{exp}$ is added to the likelihood ratio to facilitate the optimization of $P$ (see Sec.~\ref{subsec:periodic_method_more} for details).

In addition to the initial source list discussed at the beginning of this section, binary sources from the 8 yr {\em{Fermi}}-LAT 4FGL gamma-ray catalog~\citep{Fermi-LAT:2019yla} are also included in this analysis. Due to the limited sensitivity of IceCube to neutrino-induced track events in the Southern Sky, only sources with a declination greater than $-5^\circ$ are selected. Since the focus is on sources with known periodicity, the next step excludes sources without a measured period, leaving 55 sources.

{\bf{\textit{Results -}}} No evidence for periodic neutrino emission is found. The most significant source is V635 Cas, which is an HMXB consisting of a Be star and a NS. The pre-trial significance is 2.6$\,\sigma$, which results in a post-trial $p$-value of 0.25. The distribution of signal-like events is shown in the top panel of  Fig.~\ref{fig:event_dist} as a function of the phase. The full results are tabulated in Table~\ref{tab:periodic_results}.

\subsection{Flare Analysis}\label{subsec:method_2}
This analysis focuses on searching for a correlation between the neutrino emission and the X-ray activity of a source. Hard X-ray light curves are used to construct the time PDF. Light curves are obtained from data reported by {\em Swift}/BAT 15-50 keV band\footnote{\url{https://swift.gsfc.nasa.gov/results/transients/index.html}}~\citep{Krimm:2013lwa} and MAXI 10-20 keV band\footnote{\url{http://maxi.riken.jp/top/slist.html}}~\citep{matsuoka2009maxi}. The X-ray light curves are binned in days, and a Bayesian block algorithm is applied to find the optimal segmentation of the data and identify flares~\citep{Scargle:2012gq}. After the light curves are divided into blocks, the flux data points inside each segment can be fitted as a constant value taking into account the uncertainty of the X-ray data. The normalized blocked light curves over  the  total  livetime then act as the time PDF, mimicking the assumed neutrino flaring. Time-related parameters introduced here are the threshold of the X-ray flux $f_{th}$ which removes irrelevant variation thus picks flares, and the potential time lag $L_t$ between the X-ray and the neutrino emission. A more detailed description can be found in Sec.~\ref{subsec:flare_method_more}. 

Starting from our initial source list, sources without available {\em Swift}/BAT or MAXI hard X-ray light curves are removed. Furthermore, the variability and excess variance (see Sec.~\ref{subsec:flare_method_more}) of the light curves are evaluated such that sources with weak or steady emission are also removed. This step is applied only to the X-ray data in the time frame overlapped by the neutrino data sample. As neutrino emission is more likely to be correlated with harder X-ray emission, which is more probable to be nonthermal, the {\em Swift}/BAT light curves are selected over MAXI light curves cases where both pass the selection criteria. After this selection, there are 102 sources from the initial source list left to be analyzed.   

{\bf{Results -}} No significant signal events are found in the flare analysis, and results of all sources in this analysis can be found in Table~\ref{tab:lc_results}. The most significant source in the flare analysis is the microquasar V404 Cyg, a low-mass BH XRB, with a post-trial $p$-value of 0.75. There are five sub-TeV neutrino events within $1.5^{\circ}$ of the source at the time of the major X-ray flare in 2015, and the best-fit threshold indicates a time duration of 11 days, as seen in the bottom panel of Fig.~\ref{fig:event_dist}. This giant X-ray flare was observed with a duration of approximately 13 days by {\em Swift}/BAT\footnote{\url{https://www.astronomerstelegram.org/?read=7755}}. 

\subsection{Time-integrated Analysis}\label{subsec:method_3}
To complement the time-dependent studies, we conduct a time-integrated search for neutrinos on four notable sources: Cyg X-3, LS 5039, LS I +61 303, and SS 433. These are sources widely discussed as potential CR accelerators~\citep[see e.g.,][]{Gaisser:1983cj,Aharonian:2005cx,Christiansen:2005gw,Torres:2006ub,Reynoso:2019vrp,Kimura:2020acy}. Time-integrated tests on these four sources use the method described in \cite{Braun:2008bg}. 

{\bf{\textit{Results -}}} In the time-integrated analysis, we did not find any significant signal. The most significant excess is found for Cyg X-3, which exhibits a post-trial $p$-value $0.036$ after considering the 4 trials in the time-integrated analysis. Within $1^\circ$ around the source location, there are 44 events, and the most energetic among them has a deposited energy about 5 TeV, leading to a soft best-fit power-law spectrum ($\gamma=3.25$). 

\section{Discussion}
In the absence of any significant signal, 90\% confidence level (C.L.) upper limits (ULs) are computed for the sources with assumed spectra, which are shown in Table~\ref{tab:periodic_results}, Table~\ref{tab:lc_results}, and Table~\ref{tab:int_results}. There is an overlap of 30 sources between the periodic and flare analyses. V635 Cas, the source with the smallest $p$-value in the periodic analysis, shows clear flaring episodes in the {\em Swift}/BAT light curves, which yields a pre-trial $p$-value of 0.16 in the flare analysis. The most significant source in the flare analysis V404 Cyg has a pre-trial $p$-value of 0.92 in the periodic analysis. In both periodic and flaring analyses, Cyg X-3 is one of the top 10 sources. On the whole, all results are consistent with the null hypothesis.    

\begin{figure}[tb!]
    \centering
    \includegraphics[width=0.98\columnwidth]{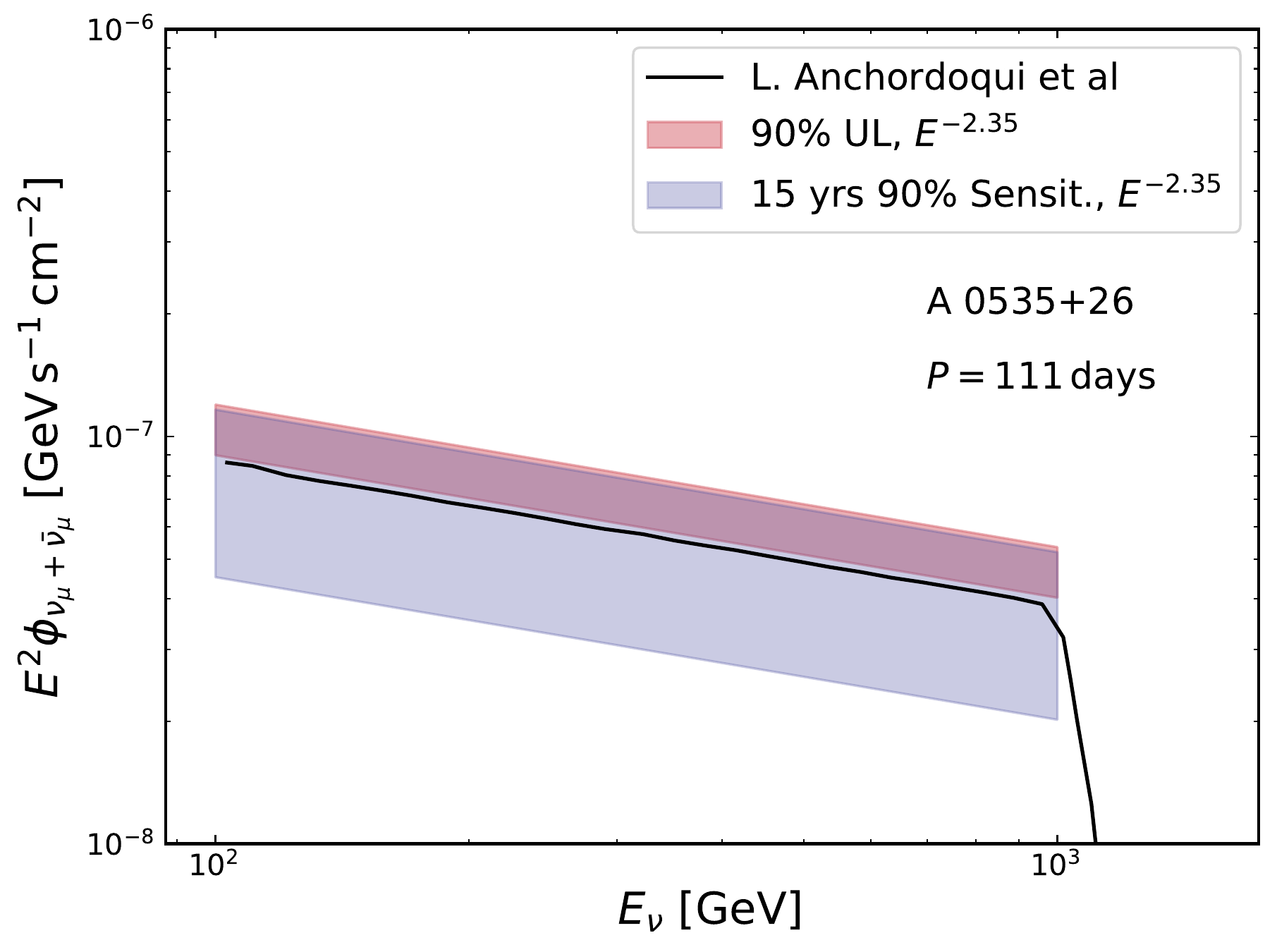}
    \caption{The neutrino flux prediction made by \cite{Anchordoqui:2002xu} is shown as the black line. The range of the UL derived from the periodic analysis under different phase concentrations is shown as the red band. The estimated sensitivity for a periodic analysis with 15 years of IceCube data is shown as the blue band.}
    \label{fig:a0535}
\end{figure}

For non-microquasar XRBs with an NS as the compact object such as V635 Cas, the most significant source in the periodic study, it is possible that the CRs are accelerated in the magnetosphere, and neutrinos can be produced via CR interactions with the accretion disk through $pp$ interactions. A similar scenario has been studied in detail for A 0535+26, an HMXB with an accreting neutron star, suggesting possible periodic neutrino emission by \cite{Anchordoqui:2002xu}. The flux prediction and the UL derived from the periodic analysis are shown in Fig.~\ref{fig:a0535}. The UL is calculated for a power-law spectrum with a cutoff at 1 TeV. The current UL is comparable to the predicted flux, yet not enough to impose a constraint on the predicted neutrino flux from the model. The range of the estimated sensitivity with 15 years of IceCube data due to varying the duty cycle is also shown in Fig.~\ref{fig:a0535}, and it suggests that this model could likely be constrained by a future IceCube analysis with an extended livetime.

\begin{figure}[tb!]
    \centering
    \includegraphics[width=1.0\columnwidth]{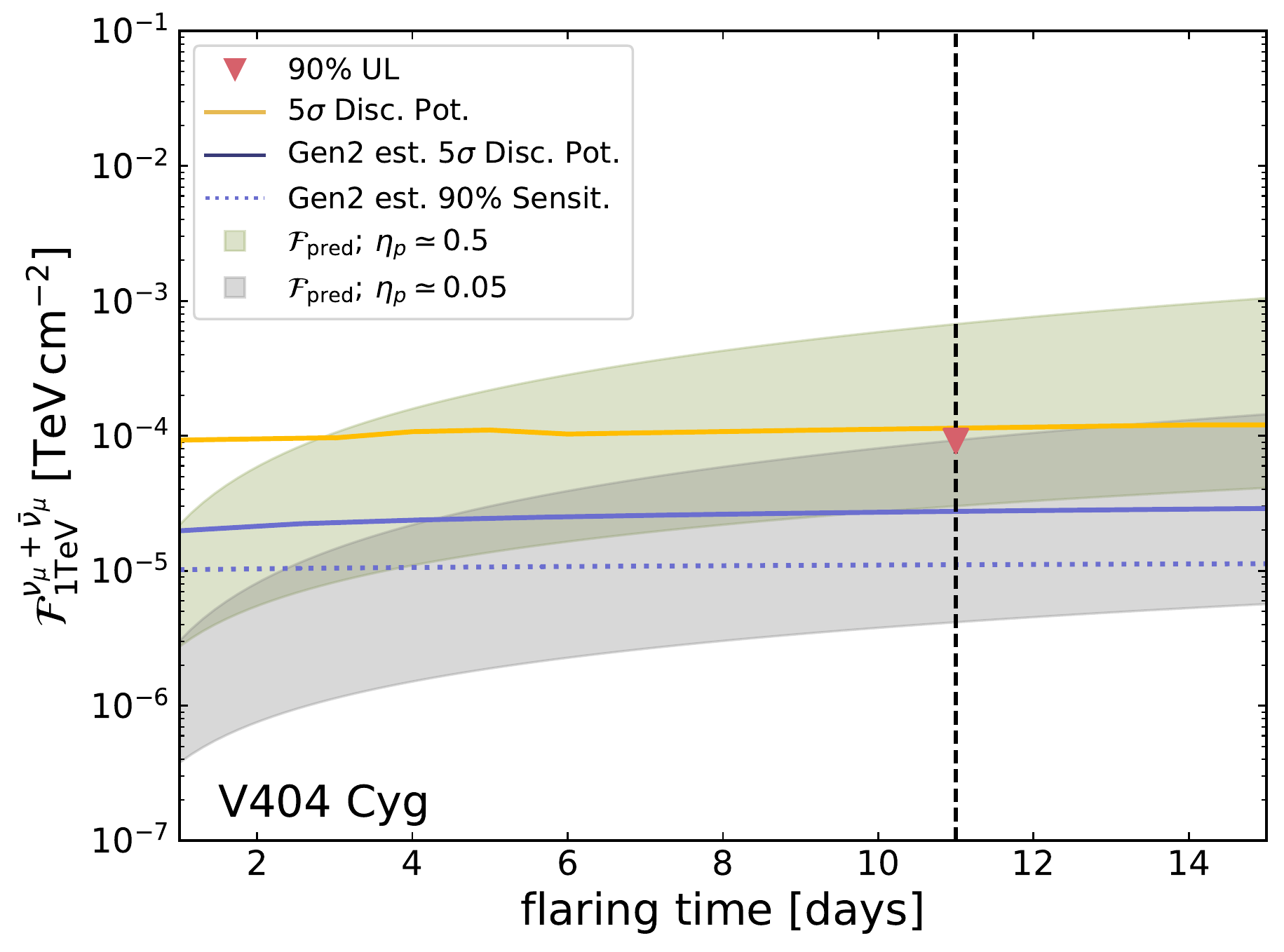} %
    \caption{The relation between the fluence at 1~TeV and the flaring time for V404 Cyg. The vertical dashed black line is the flaring time converted from the best-fit threshold, and the red triangle shows the 90\% C.L. UL in the flare search. The orange line is the 5$\sigma$ discovery potential in IceCube. Purple lines illustrate the estimated sensitivity (dotted) at 90\% C.L. and 5$\sigma$ discovery potential (solid) in IceCube-Gen2. The shaded regions are the time-integrated neutrino flux prediction estimated following the jet model~\citep{Distefano:2002qw}, assuming an $E^{-2}$ spectrum with an energy cutoff at 100~TeV. The two colors correspond to different energy fractions of the jet carried by accelerated protons $\eta_p$. The uncertainties are from flux densities in different frequencies in Very Large Array (VLA) radio measurements~\citep{Tetarenko:2018yrv} during the 2015 flare. }
    \label{fig:V404}
\end{figure}

The most significant source in the flare search, V404 Cyg, is a microquasar. For microquasars, relativistic jets are expected to be CR acceleration sites. Possible neutrino emission is expected from the beam dump on either radiation from the compact object itself or gas from the companion star. Here, we employ the parameters for neutrino flux prediction in \cite{Distefano:2002qw}, based on the photohadronic model of \cite{Levinson:2001as}. The flare of V404 Cyg in June 2015 was observed in a multi-wavelength campaign, and the jet activity during that outburst was studied, e.g., in \cite{Miller-Jones:2019zla,Tetarenko:2018yrv}. A simple estimation of the neutrino flux using the jet model can be performed with the radio jet information when the source is in an outburst state. The comparison between the ULs and the predicted fluence is shown in Fig.~\ref{fig:V404}. The CR interaction region in the source is estimated from the flaring duration, and the spectrum is assumed to follow a power-law distribution ($\gamma=2$) with an exponential cutoff at 100~TeV, motivated by spectrum of Fermi acceleration and maximal expected Galactic CR energies. Regarding jet models, the magnetic energy and the kinetic energy in the jets are usually assumed to be in equipartition, leading to large magnetic fields, which may imply a strong attenuation of pions and muons at high energies, decreasing the neutrino flux compared to the simple calculations~\citep{Reynoso:2008gs}. An alternative scenario considers neutrino emission through jet-wind interactions further away from the jet base~\citep{Reynoso:2008gs}.     

\begin{figure}[t!!]
    \centering
    \includegraphics[width=1.0\columnwidth]{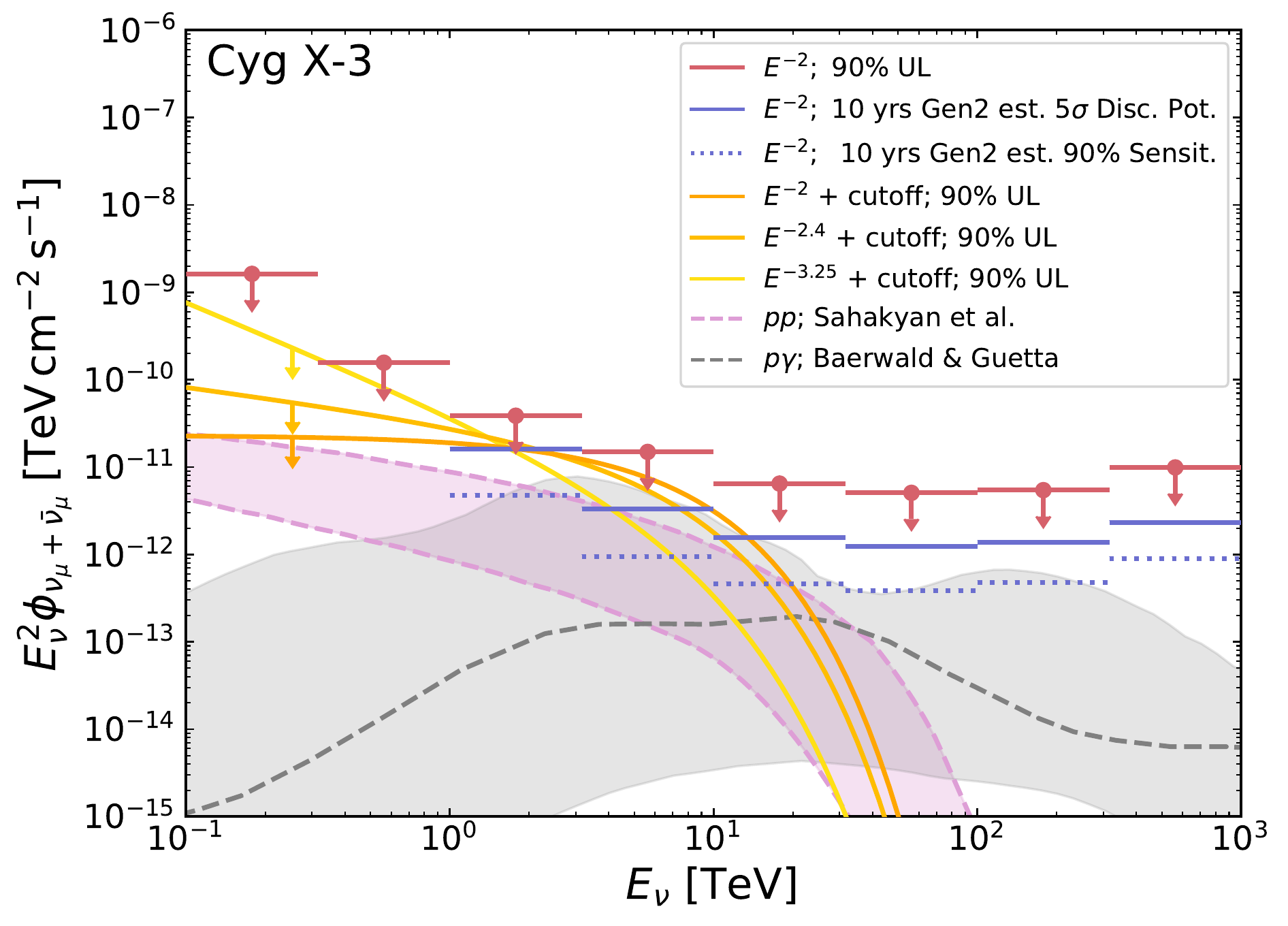} 
    \caption{Red and purple lines indicate a comparison between current UL and estimated 10 yr sensitivity (dotted) \& discovery potential (solid) in IceCube-Gen2. As the nearby events cut at several TeV, an exponential cutoff at 5 TeV is also applied for computing ULs. The shaded regions show predictions of $pp$ and $p\gamma$ scenarios. The inclusion of a cutoff is also to be compared to the shaded pink region, which includes a cutoff of CR energy at 100~TeV with the spectral index ranging from 2.4-2.7. The gray shaded region shows the uncertainty from the collision radius.}
    \label{fig:CygX-3}
\end{figure} 

Cyg X-3, the most significant source in the time-integrated search, is one of the microquasars identified as a gamma-ray source early in observations. Many predictions have been calculated in the past decades assuming different models for high-energy emission from microquasars~\citep[see e.g.,][]{Gaisser:1983cj,Kolb:1985bb,berezinskii1986high,Bednarek:2005gf}. For comparison, we take \cite{Baerwald:2012yd} and \cite{Sahakyan:2013opa}, which discussed the general $p\gamma$ and $pp$ scenarios. It is important to mention that Cyg X-3 lies in the direction of the Cygnus X region and is close to the Cygnus OB2 association, which is potentially one of the PeV point sources detected by LHAASO~\citep{cao2021ultrahigh}. Source contamination from the Cygnus X complex cannot be excluded.  
 
For TeV XRBs, in a recent ANTARES time-integrated point-source search~\citep{Illuminati:2021nvz}, HESS J0632+057 has a pre-trial $p$-value 0.02 while this periodic analysis finds a pre-trial $p$-value $9.0\times 10^{-3}$, the second most significant in the analysis. However, the flare search gives a pre-trial $p$-value of 1 while it is not included in the time-integrated search. Nevertheless, the hadronic component of the TeV gamma-ray observation cannot be constrained, and the significance is not large enough for any conclusion. For SS~433, more years of observation are needed to constrain the hadronic fraction of the observed TeV emission by HAWC~\cite{HAWC:2018gwz} and flux prediction in e.g.,~\cite{Kimura:2020acy} while prediction in~\cite{Distefano:2002qw} is strongly constrained. More data are also needed to constrain LS 5039, which lies in the Southern Sky, and LS~I~+61~303, where the neutrino flux calculation from e.g.,~\cite{Torres:2006ub}  is not constrained by the upper limit.

The next generation of the IceCube Observatory, IceCube-Gen2, will provide a factor of eight extension in volume~\citep{Aartsen:2020fgd}, leading to an expected five-fold increase in the effective area compared to IceCube, corresponding to an improvement in sensitivity by the same order. Here, we extend the study to IceCube-Gen2 and estimate the sensitivity and discovery potential for V404 Cyg, as an example of a flaring source and Cyg X-3, for persistently emitting sources. The estimated improvement can be seen in Fig.~\ref{fig:V404} and Fig.~\ref{fig:CygX-3}. Here, the effective areas of muon tracks are computed from the proposed IceCube-Gen2 configuration, and the projection is evaluated with the same method as in \cite{Aartsen:2020fgd} without considering contribution from the existing IceCube detector. In comparison with theoretical calculations, it demonstrates the potential to either identify those sources or better constrain models in the future. Therefore, providing a better estimation of the hadronic component of TeV gamma-ray sources.

\section{Conclusion \& Outlook}

In this paper, we presented a comprehensive study on neutrino emission from XRBs, a long-standing candidate for the Galactic sources of CRs and neutrinos. With no significant signal events found, we set ULs on the neutrino emission in the scenarios presented with a discussion. Our estimation with the improved sensitivity of IceCube-Gen2 demonstrates the potential of future detection, presenting a promising outlook of identifying XRBs as Galactic cosmic-ray accelerators in the upcoming years.

\section*{Acknowledgements }
The IceCube collaboration acknowledges the significant contributions to this manuscript from Ali Kheirandish, Qinrui Liu and Chun Fai Tung. The authors gratefully acknowledge the support from the following agencies and institutions: USA {\textendash} U.S. National Science Foundation-Office of Polar Programs,
U.S. National Science Foundation-Physics Division,
U.S. National Science Foundation-EPSCoR,
Wisconsin Alumni Research Foundation,
Center for High Throughput Computing (CHTC) at the University of Wisconsin{\textendash}Madison,
Open Science Grid (OSG),
Extreme Science and Engineering Discovery Environment (XSEDE),
Frontera computing project at the Texas Advanced Computing Center,
U.S. Department of Energy-National Energy Research Scientific Computing Center,
Particle astrophysics research computing center at the University of Maryland,
Institute for Cyber-Enabled Research at Michigan State University,
and Astroparticle physics computational facility at Marquette University;
Belgium {\textendash} Funds for Scientific Research (FRS-FNRS and FWO),
FWO Odysseus and Big Science programmes,
and Belgian Federal Science Policy Office (Belspo);
Germany {\textendash} Bundesministerium f{\"u}r Bildung und Forschung (BMBF),
Deutsche Forschungsgemeinschaft (DFG),
Helmholtz Alliance for Astroparticle Physics (HAP),
Initiative and Networking Fund of the Helmholtz Association,
Deutsches Elektronen Synchrotron (DESY),
and High Performance Computing cluster of the RWTH Aachen;
Sweden {\textendash} Swedish Research Council,
Swedish Polar Research Secretariat,
Swedish National Infrastructure for Computing (SNIC),
and Knut and Alice Wallenberg Foundation;
Australia {\textendash} Australian Research Council;
Canada {\textendash} Natural Sciences and Engineering Research Council of Canada,
Calcul Qu{\'e}bec, Compute Ontario, Canada Foundation for Innovation, WestGrid, and Compute Canada;
Denmark {\textendash} Villum Fonden and Carlsberg Foundation;
New Zealand {\textendash} Marsden Fund;
Japan {\textendash} Japan Society for Promotion of Science (JSPS)
and Institute for Global Prominent Research (IGPR) of Chiba University;
Korea {\textendash} National Research Foundation of Korea (NRF);
Switzerland {\textendash} Swiss National Science Foundation (SNSF);
United Kingdom {\textendash} Department of Physics, University of Oxford.\\

\bibliography{bibfile}

\appendix

\section{Method}\label{sec:method_details}
The general form of the likelihood in this work can be written as 

\begin{equation}\label{eq:generic_llh}
    \mathcal{L} = \prod_i^{N} \left[ \frac{n_s}{N} \mathcal{S}_i + \left( 1 - \frac{n_s}{N} \right) \mathcal{B}_i \right],
\end{equation}
where $i$ indicates the i-th event, $n_s$ is the number of signal events, $N$ is the total number of events, and $\mathcal{S}_i$, $\mathcal{B}_i$ are the signal and background PDFs evaluated for event $i$, respectively. The signal PDF is the product of three terms,
\begin{equation}\label{eq:pdf_sig}
    \mathcal{S}_i = \mathcal{P}\left(x_i, \sigma_i | x_s \right) \times \mathcal{E}\left(E_i | \gamma \right) \times \mathcal{T}\left( t_i | \theta_1, \theta_2...\right),
\end{equation}
where $\mathcal{P}, \mathcal{E}$ and $\mathcal{T}$ are the PDFs of space, energy and time, respectively. $x_i$ is the reconstructed incoming direction, $\sigma_i$ is the estimated angular uncertainty, $E_i$ is the reconstructed muon energy, and $t_i$ is the arrival time of the event. The spatial clustering of signal events, represented by $\mathcal{P}$, can be modeled as a 2D Gaussian distribution; the energy term $\mathcal{E}$ is modeled by a simple power-law spectrum with spectral index $\gamma$, and the detector response is determined by the Monte Carlo simulation. In this work, the fitting range of $\gamma$ is set to 1-4. For a time-integrated search, the time term $\mathcal{T}$ corresponds to a uniform distribution, in search for a steady emission of neutrinos. For sources with flaring or periodic characteristics, signal events can cluster in small timescales. Inclusion of a time selection can reduce the background effectively, thus improve the sensitivity~\citep{Braun:2009wp}. The background PDF $\mathcal{B}_i=1/(2\pi)\mathcal{P}_B(\delta_i)\times \mathcal{E}_B(E_i)$ is constructed by binning the experimental data in reconstructed declination and energy, based on the assumption that background events dominate the data. It is independent of the right ascension due to the Earth's rotation. The time distribution of the background is assumed to be uniform, considering that the effect of seasonal variation is negligible.

\subsection{Periodic Search}\label{subsec:periodic_method_more}
The signal time PDF is expressed as a phase PDF modeled by the von Mises distribution
\begin{equation}
    \mathcal{T}(t_i|\kappa, \Phi_0, P) = \frac{1}{2 \pi I_0\left(\kappa\right)}\exp{ \left[ \kappa\cos\left(\phi_i \left( t_i| P \right) - \Phi_0 \right) \right]},
\end{equation}
where $\phi_i$ is the phase of the event which is a function of the arrival time, $t_i$ and the orbital period, $P$ of the source using the equation $\phi_i = (t_i-T_0)/P$, with $T_0$ set to MJD 55690. The value $T_0$ is chosen to reduce the error accumulated by the possible uncertainty in $P$.  $\kappa$ measures the concentration of events within a period, and $\Phi_0$ is the mean phase of the distribution. $\kappa$, $\Phi_0$, $P$, along with $n_s$ and $\gamma$, are free parameters.

The TS in this analysis is 
\begin{equation}
        TS = -2\log \left[ \frac{\mathcal{L} \left( n_s = 0 \right)}{\mathcal{L} \left( \hat{n}_s, \hat{\gamma}, \hat{\kappa}, \hat{\Phi}_0, \hat{P} \right)} \times \frac{2\pi}{\sigma_{\Psi}(\hat{\kappa})} \right]-2\frac{(\hat{P}-P_{\rm exp})^2}{2\sigma_P^2}.
\end{equation}

Similar to the previous time-dependent analysis by \cite{Aartsen:2015wto}, the likelihood ratio is modified by a marginalization term $2\pi/\sigma_{\Psi}(\hat{\kappa})$ to prevent the minimizer from showing biases towards short flares, where $\sigma_{\Psi}$ is the standard deviation of the von Mises distribution given by
\begin{equation}
    \sigma_{\Psi}(\kappa) = \sqrt{-2\log\left( \frac{I_1(\kappa)}{I_0(\kappa)} \right)}.
\end{equation}
The second term is the prior probability of the orbital period of the source. It is approximated by a normal distribution centered at the measured orbital period $P_{exp}$, and the variance $\sigma^2_P$ is chosen as the reported uncertainty of $P_{\rm exp}$. The role of the prior is to facilitate the minimization of the orbital period with the information provided by the electromagnetic observations of the sources.

\subsection{Flare Search}\label{subsec:flare_method_more}
The method is similar to the IceCube triggered flare search in~\cite{Aartsen:2015wto}. The Bayesian block algorithm is applied to the daily binned X-ray data to characterize statistically significant variations, thus remove noise and identify flares. The signal time PDF is constructed from the Bayesian blocked X-ray light curve and can be written as

\begin{equation}
    \mathcal{T}(t_i|f_{\rm th},t_{\rm lag}) = \frac{max\left (0,\;f(t_i-t_{\rm lag})-f_{\rm th} \right) }{\int_{T_{\rm min}}^{T_{\rm max}} max\left (0,\;f(t_i-t_{\rm lag})-f_{\rm th}\right )\,dt},
\end{equation}
where $f$ is the count rate of the light curve, which is a function of time and $f_{th}$ is the count rate threshold. Only the light curve above the threshold is kept, which picks out the flares we care about while the irrelevant variation below the threshold is removed. $t_{\rm lag}$ is the time lag, which accounts for the time difference between X-ray flares and neutrino flares. This time lag comes from two factors: the first is the time difference between X-ray and neutrino emissions, which is not expected to be a large value; the second is the uncertainty of binning, as X-ray light curves used in this analysis are daily binned and the real emission time is unknown. We allow a uniform time lag prior of $\pm 0.5$ days for fitting. The TS is defined as the maximal log-likelihood ratio $\rm{TS} = -2\,{\log}[ (\mathcal{L}(n_s=0)/\mathcal{L}(\hat{n}_s,\hat{\gamma},\hat{f}_{\rm th},\hat{t}_{\rm lag})]$.

The source selection is based on the variability which describes how variable the data is, and the excess variance which identifies flares of X-ray light curves~\citep{Krimm:2013lwa}

\begin{equation}
\begin{split}
    V_X &=\frac{1}{N_X-1}\sum_i^{N_X}\frac{(F_{X,\,i}-F_{X,\,\rm{avg}})^2}{\sigma_{X,i}}\\
    F_{X,\rm{var}}& = \frac{1}{F_{X,\,\rm{avg}}}\sqrt{\frac{\sum (F_{X,\,i}-F_{X,\,\rm{avg}})^2}{N_X-1}-\frac{\sum \sigma^2_{X,i}}{N}},
    \end{split}
\end{equation}
where $F_{X,\,i}$ is the individual measurement of a source flux, $F_{X,\,\rm{avg}}$ is the weighted average and $N_X$ is the total number of measurements. $\sigma_{X,\, i}$ includes both statistical and systematic errors. 

\section{source tables}\label{source_tables}
The initial source list includes 301 X-ray binary sources from the HMXB and LMXB catalogs. On top of those, 7 binary sources from the TeVCat, which are not listed in the previous two, are added to the initial list. In the periodic search, 31 sources from the HMXB, 22 from the LMXB, and 1 from TeVCat HESS J0632+057, which have resolved orbital periods and declinations above $-5^\circ$ are selected. An extra source, 3FGL J0212.1+5320 from the 4FGL catalog, is added to form the analysis source list of 55 sources.
In the flare search, there are 43 HMXB and 58 LMXB from the catalogs covering the whole sky with one extra HMXB HESS J0632+057 from TeVCat, which compose the total 102 sources after the selection. There are 30 sources overlapping in the two analyses.  
\centerwidetable 
\startlongtable
\begin{deluxetable*}{lDDDDDDDDDDD}\movetableright=-1in
\tablecaption{Sources and Results of the Periodic Search}
\tablehead{
\colhead{Source} & \multicolumn2c{RA [deg]} & \multicolumn2c{$\delta$ [deg]} & \multicolumn2c{TS} & \multicolumn2c{$\hat{n}_{s}$} & \multicolumn2c{$\hat{\gamma}$} & \multicolumn2c{$\hat{\kappa}$} & \multicolumn2c{$\hat{\Phi}_0$} & \multicolumn2c{$\hat{P}$} & \multicolumn2c{p-value} & \multicolumn2c{$\Phi_{\nu_{\mu}+\bar{\nu}_{\mu}}^{90\%,\;E^{-2.0}}$} & \multicolumn2c{$\Phi_{\nu_{\mu}+\bar{\nu}_{\mu}}^{90\%,\;E^{-3.0}}$}
}
\decimals
\startdata
V1037 Cas & 7.26 & 59.57 & -2.63 & 0.0 & 3.92 & 0.5 & 1.63 & 0.102 & 0.76 & 0.74 & 8.17 \\
BD +60 73 & 9.29 & 61.36 & -2.36 & 6.9 & 4 & 0.5 & 5.8 & 15.66 & 0.51 & 0.71 & 8.3 \\
4U 0042+32 & 11.21 & 33.02 & 2.56 & 17.5 & 4 & 21.1 & 4.12 & 11.588 & 0.22 & 0.77 & 16.1 \\
gam Cas & 14.18 & 60.72 & -1.79 & 13.4 & 4 & 0.5 & 1.68 & 203.59 & 0.41 & 0.78 & 9.11 \\
V662 Cas & 19.51 & 65.29 & 0.1 & 25.8 & 4 & 1.79 & 4.02 & 11.599 & 0.22 & 1.29 & 14.2 \\
V635 Cas & 19.63 & 63.74 & 9.07 & 50.5 & 4 & 0.83 & 2.58 & 24.316 & 0.0052 & 2.67 & 28.5 \\
3FGL J0212.1+5320 & 33.04 & 53.35 & 1.04 & 19.8 & 3.01 & 0.5 & 3.38 & 0.869 & 0.19 & 1.08 & 14.6 \\
LS I +61 303 & 40.13 & 61.23 & -2.63 & 0.0 & 3.54 & 0.5 & 2.18 & 26.496 & 0.86 & 0.71 & 7.99 \\
BQ Cam & 53.75 & 53.17 & 0.108 & 17.4 & 3.3 & 10.2 & 1.34 & 33.85 & 0.17 & 0.93 & 14.6 \\
X Per & 58.85 & 31.05 & -1.17 & 15.6 & 3.89 & 2.26 & 4.63 & 250.3 & 0.31 & 0.62 & 11.4 \\
CI Cam & 64.93 & 56.00 & -2.63 & 0.0 & 3.99 & 0.5 & 4.99 & 19.41 & 0.89 & 0.72 & 7.74 \\
V518 Per & 65.43 & 32.91 & -2.16 & 9.4 & 3.48 & 0.5 & 2.54 & 0.212 & 0.48 & 0.55 & 8.18 \\
LS V +44 17 & 70.25 & 44.53 & -2.63 & 0.0 & 3.14 & 0.5 & 1.78 & 150 & 0.90 & 0.55 & 7.75 \\
1A 0535+262 & 84.73 & 26.32 & -2.63 & 0.0 & 3.96 & 0.5 & 0.08 & 111 & 0.70 & 0.47 & 8.93 \\
Swift J061223.0+701243.9 & 93.09 & 70.21 & -0.439 & 1.6 & 1.19 & 0.5 & 0.84 & 0.013 & 0.32 & 1.34 & 13.6 \\
V1055 Ori & 94.28 & 9.14 & 4.75 & 17.8 & 2.65 & 59.6 & 6.18 & 0.028 & 0.099 & 0.69 & 25.5 \\
V616 Mon & 95.69 & -0.35 & 0.523 & 27.5 & 3.71 & 1.26 & 4.7 & 0.323 & 0.20 & 0.44 & 18.9 \\
HESS J0632+057 & 98.25 & 5.80 & 11.3 & 39.7 & 3.73 & 23.6 & 0.75 & 308.75 & 0.0090 & 0.78 & 38.5 \\
SAX J0635.2+0533 & 98.83 & 5.55 & 0.308 & 18.8 & 2.93 & 10.1 & 0.48 & 11.168 & 0.36 & 0.38 & 14.7 \\
KV UMa & 169.55 & 48.04 & -2.23 & 1.6 & 1.99 & 0.5 & 3 & 0.170 & 0.51 & 0.58 & 7.86 \\
UW CrB & 241.44 & 25.86 & -1.39 & 17.8 & 4 & 0.5 & 5.92 & 0.077 & 0.34 & 0.63 & 12.2 \\
Her X-1 & 254.46 & 35.34 & -0.718 & 20.1 & 4 & 0.5 & 2.08 & 1.700 & 0.26 & 0.75 & 12.1 \\
V934 Her & 256.64 & 23.97 & -2.63 & 1.2 & 3.98 & 0.5 & 1.24 & 420.17 & 0.72 & 0.40 & 7.8 \\
Swift J1753.5-0127 & 268.37 & -1.45 & -2.63 & 0.0 & 2.61 & 0.5 & 2.74 & 0.135 & 0.73 & 0.25 & 9.95 \\
4U 1823-00 & 276.34 & -0.01 & 3.87 & 22.7 & 3.38 & 18.6 & 5.35 & 0.133 & 0.16 & 0.43 & 23.6 \\
AX 1845.0-0433 & 281.26 & -4.56 & 2.42 & 30.6 & 3.93 & 0.5 & 5.15 & 5.720 & 0.056 & 0.69 & 35 \\
IGR J18483-0311 & 282.07 & -3.18 & -2.63 & 0.0 & 3.16 & 0.5 & 1.32 & 18.545 & 0.65 & 0.27 & 11.1 \\
2S 1845-024 & 282.07 & -2.42 & -2.63 & 0.0 & 2.76 & 0.5 & 4.81 & 242.18 & 0.95 & 0.24 & 11.4 \\
XTE J1855-026 & 283.88 & -2.61 & -2.54 & 4.7 & 3.72 & 0.5 & 4.34 & 6.074 & 0.53 & 0.28 & 10.8 \\
V406 Vul & 284.67 & 22.66 & 1.26 & 23.9 & 3.34 & 5.06 & 0.89 & 0.274 & 0.27 & 0.60 & 14.6 \\
XTE J1859+083 & 284.77 & 8.25 & -2.63 & 0.0 & 2.16 & 0.5 & 4.29 & 60.65 & 0.87 & 0.31 & 9.75 \\
4U 1901+03 & 285.90 & 3.19 & -2.63 & 0.0 & 3.46 & 0.5 & 4.65 & 22.535 & 0.67 & 0.28 & 10.8 \\
4U 1907+09 & 287.41 & 9.83 & 2.7 & 37.6 & 4 & 0.5 & 2.97 & 8.375 & 0.060 & 0.69 & 24.7 \\
4U 1909+07 & 287.70 & 7.60 & 0.14 & 29.0 & 4 & 0.5 & 4.93 & 4.400 & 0.18 & 0.58 & 17.6 \\
Aql X-1 & 287.82 & 0.58 & -2.53 & 2.0 & 2.48 & 0.5 & 4.6 & 0.790 & 0.64 & 0.26 & 9.9 \\
SS 433 & 287.96 & 4.98 & -2.63 & 0.0 & 2.76 & 0.5 & 3.26 & 13.068 & 0.84 & 0.25 & 9.95 \\
IGR J19140+0951 & 288.52 & 9.88 & -0.162 & 24.6 & 4 & 6.27 & 2.99 & 13.553 & 0.20 & 0.47 & 16.9 \\
GRS 1915+105 & 288.80 & 10.95 & -0.865 & 23.4 & 3.77 & 0.5 & 2.78 & 33.862 & 0.31 & 0.52 & 15 \\
XTE J1946+274 & 296.41 & 27.37 & 5.61 & 35.8 & 3.52 & 2.87 & 5.58 & 171.1 & 0.026 & 0.11 & 24.7 \\
KS 1947+300 & 297.38 & 30.21 & -2.63 & 0.0 & 3.03 & 0.5 & 1.32 & 40.415 & 0.91 & 0.47 & 8.29 \\
Cyg X-1 & 299.59 & 35.20 & -2.63 & 0.0 & 1.25 & 0.5 & 0.39 & 5.600 & 0.83 & 0.50 & 7.89 \\
V1408 Aql & 299.85 & 11.71 & -2.63 & 0.0 & 2.77 & 0.5 & 1.34 & 0.389 & 0.91 & 0.34 & 8.92 \\
QZ Vul & 300.71 & 25.24 & -2.63 & 0.0 & 2.64 & 0.5 & 5.01 & 0.344 & 0.78 & 0.48 & 8.38 \\
V404 Cyg & 306.02 & 33.87 & -2.63 & 0.0 & 4 & 0.5 & 1.6 & 6.473 & 0.92 & 0.47 & 7.89 \\
V2246 Cyg & 308.06 & 37.64 & -0.213 & 1.5 & 1.17 & 0.5 & 3.06 & 46.021 & 0.19 & 0.84 & 13.2 \\
Cyg X-3 & 308.11 & 40.95 & 7.46 & 50.6 & 3.52 & 1.06 & 4.82 & 0.200 & 0.045 & 1.38 & 22.9 \\
GRO J2058+42 & 314.70 & 41.78 & -2.63 & 0.0 & 3.88 & 0.5 & 0.65 & 55.03 & 0.79 & 0.51 & 7.54 \\
SAX J2103.5+4545 & 315.90 & 45.75 & -2.46 & 4.0 & 2.55 & 0.5 & 1.7 & 12.665 & 0.52 & 0.53 & 7.9 \\
M15 X-2 & 322.49 & 12.17 & -2.37 & 8.2 & 3.8 & 0.5 & 6.27 & 0.016 & 0.56 & 0.35 & 10 \\
M15 X-1 & 322.49 & 12.17 & -2.59 & 3.5 & 3.99 & 0.5 & 3.01 & 0.713 & 0.64 & 0.30 & 9.43 \\
V1727 Cyg & 322.86 & 47.29 & 0.452 & 17.4 & 4 & 8.35 & 0.15 & 0.218 & 0.16 & 0.91 & 14.7 \\
V490 Cep & 324.88 & 56.99 & -0.477 & 1.5 & 1 & 1.33 & 5.31 & 20.848 & 0.32 & 0.90 & 12 \\
Cyg X-2 & 326.17 & 38.32 & -0.0499 & 20.6 & 4 & 0.5 & 5.95 & 9.845 & 0.20 & 0.90 & 13.3 \\
BD +53 2790 & 331.98 & 54.52 & -0.275 & 16.3 & 4 & 1.98 & 1.02 & 9.56 & 0.21 & 1.07 & 12.8 \\
SAX J2239.3+6116 & 339.84 & 61.27 & 3.09 & 23.5 & 4 & 12.1 & 0.6 & 265.62 & 0.088 & 1.22 & 16.4 \\
\enddata
\tablecomments{Sources for the periodic point-source analysis with their best-fitted TS, $\hat{n}_s$, $\hat{\gamma}$, $\hat{\kappa}$, $\hat{\Phi}_0$, $P$ (in days), and pre-trial $p$-values. The 90\% C.L. flux ULs are parameterized as $dN_{\nu_\mu+\bar{\nu}_\mu}/dE_\nu=\Phi_{\nu_\mu+\bar{\nu}}^{90\%,\;E^{-\gamma}}\left ({E_\nu}/{\rm{TeV}} \right )^{-\gamma}\;$ where $\gamma$ indicates the assumed spectral index of a power-law spectrum, and their units are $10^{-12}\;\rm{TeV^{-1}cm^{-1}s^{-1}}$. The equatorial coordinates are provided in the J2000 epoch, which also applies to other tables in this paper.} 
\label{tab:periodic_results}
\end{deluxetable*}

\centerwidetable 
\startlongtable
\begin{deluxetable*}{lDDDDDDDDDDD}\movetableright=-1in
\tablecaption{Sources and Results of the Flare Search}
\tablehead{
\colhead{Source} & \multicolumn2c{RA [deg]} & \multicolumn2c{$\delta$ [deg]} & \multicolumn2c{TS} & \multicolumn2c{$\hat{n}_{s}$} & \multicolumn2c{$\hat{\gamma}$} & \multicolumn2c{$\hat{f}_{th}$} & \multicolumn2c{$\hat{T}_{lag}$} & \multicolumn2c{$p$-value} & \multicolumn2c{$\mathcal{F}_{\nu_{\mu}+\bar{\nu}_{\mu}}^{90\%,\;E^{-2.0}}$} & \multicolumn2c{$\mathcal{F}_{\nu_{\mu}+\bar{\nu}_{\mu}}^{90\%,\;E^{-2.5}}$} & \multicolumn2c{$\mathcal{F}_{\nu_{\mu}+\bar{\nu}_{\mu}}^{90\%,\;E^{-3.0}}$}
}
\decimals
\startdata
V662 Cas & 19.51 & 65.29 & 1.54 & 14.8 & 4 & 0.03 & -0.16 & 0.54 & 0.15 & 0.5 & 1.1 \\
V635 Cas & 19.63 & 63.74 & 2.69 & 2.3 & 2.46 & 8.23 & -0.46 & 0.16 & 0.11 & 0.3 & 0.5 \\
RX J0146.9+6121 & 26.75 & 61.36 & 0 & 0 & - & - & - & 1 & 0.30 & 0.3 & 2.6 \\
BQ Cam & 53.75 & 53.17 & 2.25 & 2.9 & 2.96 & 18.55 & -0.25 & 0.16 & 0.09 & 0.4 & 0.6 \\
X Per & 58.85 & 31.05 & 2.62 & 17.2 & 3.60 & 0.60 & 0.49 & 0.37 & 0.07 & 0.4 & 0.9 \\
LS V +44 17 & 70.25 & 44.53 & 1.53 & 1.9 & 4 & 0.58 & -0.03 & 0.24 & 0.07 & 0.2 & 0.5 \\
1A 0535+262 & 84.73 & 26.32 & 0 & 0 & - & - & - & 1 & 0.15 & 0.2 & 1.6 \\
V1055 Ori & 94.28 & 9.14 & 1.75 & 11.7 & 2.56 & 0.00 & 0.42 & 0.42 & 0.09 & 0.7 & 1.9 \\
HESS J0632+057 & 98.25 & 5.80 & 0 & 0 & - & - & - & 1 & 0.14 & 0.3 & 3.1 \\
GS 0834-430 & 128.98 & -43.19 & 0 & 0 & - & - & - & 1 & 1.97 & 22.3 & 1536.4 \\
Vela X-1 & 135.53 & -40.55 & 2.22 & 2.8 & 4 & 6.73 & 0.35 & 0.41 & 0.95 & 33.9 & 704.2 \\
GRO J1008-57 & 152.45 & -58.29 & 0 & 0 & - & - & - & 1 & 3.76 & 38.2 & 3021.9 \\
Cen X-3 & 170.31 & -60.62 & 4.86 & 1.1 & 1.00 & 2.49 & -0.05 & 0.18 & 1.83 & 78.3 & 1647.4 \\
1E 1145.1-6141 & 176.87 & -61.95 & 7.54 & 10.4 & 2.33 & 0.18 & 0.11 & 0.05 & 2.65 & 107.1 & 2295.8 \\
GX 301-2 & 186.66 & -62.77 & 2.43 & 1.8 & 2.49 & 18.91 & -0.44 & 0.33 & 0.93 & 32.7 & 708.5 \\
1A 1246-588 & 192.40 & -59.12 & 0.20 & 0.5 & 2.00 & 4.47 $^m$ & -0.00 & 0.63 & 0.72 & 26.1 & 512.3 \\
GX 304-1 & 195.32 & -61.60 & 0.07 & 0.3 & 2.87 & 1.57 & -0.16 & 0.47 & 0.97 & 32.8 & 612.0 \\
4U 1323-619 & 201.65 & -62.14 & 1.57 & 0.8 & 4 & 5.36 $^m$ & -0.50 & 0.38 & 0.80 & 26.6 & 520.8 \\
Ginga 1354-645 & 209.54 & -64.73 & 0 & 0 & - & - & - & 1 & 3.13 & 35.3 & 2371.7 \\
H 1417-624 & 215.30 & -62.70 & 0.13 & 0.5 & 1.60 & 0.12 & -0.31 & 0.40 & 0.94 & 32.0 & 755.9 \\
Cir X-1 & 230.17 & -57.17 & 3.59 & 1.0 & 1.64 & 1.12 & 0.50 & 0.13 & 1.10 & 34.5 & 761.4 \\
H 1538-522 & 235.60 & -52.39 & 3.35 & 2.1 & 4 & 0.68 & 0.50 & 0.22 & 1.26 & 40.0 & 899.5 \\
H 1553-542 & 239.45 & -54.41 & 0 & 0 & - & - & - & 1 & 2.74 & 29.3 & 2325.7 \\
H 1608-522 & 243.18 & -52.42 & 0.13 & 0.4 & 4 & 1.23 & 0.50 & 0.59 & 0.71 & 27.5 & 532.0 \\
Sco X-1 & 244.98 & -15.64 & 3.86 & 6.8 & 4 & 29.05 & 0.33 & 0.18 & 0.40 & 12.6 & 202.0 \\
IGR J16318-4848 & 247.95 & -48.82 & 5.75 & 5.4 & 3.59 & 0.42 & 0.38 & 0.12 & 1.69 & 65.2 & 1476.7 \\
AX J1631.9-4752 & 248.01 & -47.87 & 2.87 & 0.9 & 1.75 & 1.80 & 0.47 & 0.34 & 0.60 & 18.5 & 386.0 \\
4U 1630-472 & 248.51 & -47.39 & 0 & 0 & - & - & - & 1 & 2.49 & 24.7 & 2214.6 \\
4U 1636-536 & 250.23 & -53.75 & 0 & 0 & - & - & - & 1 & 5.96 & 63.8 & 5666.3 \\
GX 340+0 & 251.45 & -45.61 & 0.54 & 1.5 & 3.13 & 1.67 & -0.18 & 0.69 & 0.69 & 21.1 & 467.1 \\
IGR J16479-4514 & 252.03 & -45.20 & 6.11 & 2.7 & 2.35 & 0.96 & -0.38 & 0.08 & 0.90 & 31.7 & 665.5 \\
Her X-1 & 254.46 & 35.34 & 0 & 0 & - & - & - & 1 & 0.41 & 0.6 & 5.9 \\
EXO 1657-419 & 255.20 & -41.66 & 8.44 & 2.1 & 3.10 & 6.70 & -0.50 & 0.02 & 1.05 & 33.2 & 690.8 \\
GX 339-4 & 255.71 & -48.79 & 0.22 & 0.8 & 1.96 & 2.67 & 0.47 & 0.54 & 0.67 & 24.5 & 497.3 \\
4U 1700-377 & 255.99 & -37.84 & 0.08 & 0.5 & 4 & 2.75 & 0.50 & 0.82 & 0.93 & 35.6 & 782.2 \\
GX 349+2 & 256.44 & -36.42 & 0.10 & 1.3 & 2.44 & 1.65 & -0.41 & 0.86 & 0.69 & 25.9 & 667.4 \\
4U 1702-429 & 256.56 & -43.04 & 0 & 0 & - & - & - & 1 & 4.36 & 45.6 & 4569.1 \\
4U 1700+24 & 256.64 & 23.97 & 3.38 & 6.5 & 4 & 0.48 & 0.44 & 0.23 & 0.07 & 0.4 & 0.8 \\
H 1705-440 & 257.23 & -44.10 & 4.60 & 6.6 & 4 & 0.48 & 0.13 & 0.09 & 1.61 & 62.5 & 1404.4 \\
IGR J17091-3624 & 257.26 & -36.39 & 1.35 & 3.0 & 2.31 & 0.32 & -0.50 & 0.30 & 0.78 & 24.3 & 520.7 \\
Granat 1716-249 & 259.90 & -25.02 & 1.24 & 2.1 & 2.97 & 11.05 & -0.21 & 0.34 & 0.38 & 12.7 & 227.6 \\
IGR J17252-3616 & 261.30 & -36.28 & 0.34 & 0.5 & 1.14 & 0.57 & -0.48 & 0.69 & 0.46 & 15.9 & 307.4 \\
4U 1722-30 & 261.89 & -30.80 & 0 & 0 & - & - & - & 1 & 2.58 & 23.5 & 2184.6 \\
GX 9+9 & 262.93 & -16.96 & 6.62 & 1.0 & 2.24 & 20.27 $^m$ & -0.47 & 0.06 & 0.35 & 8.3 & 111.4 \\
GX 354-0 & 262.99 & -33.83 & 0 & 0 & - & - & - & 1 & 3.23 & 34.7 & 3641.9 \\
GX 1+4 & 263.01 & -24.75 & 0.13 & 0.5 & 3.38 & 2.52 & -0.50 & 0.71 & 0.38 & 11.8 & 240.5 \\
Rapid Burster & 263.35 & -33.39 & 0 & 0 & - & - & - & 1 & 2.40 & 21.6 & 2043.3 \\
SLX 1732-304 & 263.95 & -30.48 & 3.43 & 7.9 & 3.87 & 0.00 $^m$ & 0.25 & 0.17 & 0.92 & 34.9 & 675.3 \\
4U 1735-44 & 264.74 & -44.45 & 0 & 0 & - & - & - & 1 & 4.00 & 36.6 & 3838.4 \\
 IGR J17391-3021 & 264.80 & -30.34 & 0.75 & 0.8 & 2.53 & 0.07 & -0.47 & 0.32 & 0.41 & 13.0 & 231.8 \\
GRS 1739-278 & 265.67 & -27.75 & 0 & 0 & - & - & - & 1 & 1.43 & 12.6 & 1131.0 \\
1E 1740.7-2942 & 265.98 & -29.75 & 0 & 0 & - & - & - & 1 & 3.05 & 33.9 & 2977.0 \\
GRO J1744-28 & 266.14 & -28.74 & 0.02 & 0.2 & 2.30 & 0.54 & 0.34 & 0.51 & 0.44 & 14.1 & 249.3 \\
AX J1744.8-2921 & 266.21 & -29.35 & 0 & 0 & - & - & - & 1 & 2.38 & 21.7 & 2297.1 \\
Granat J1741.9-2853 & 266.26 & -28.91 & 0 & 0 & - & - & - & 1 & 1.45 & 12.1 & 836.0 \\
AX J1745.6-2901 & 266.42 & -29.01 & 1.50 & 1.1 & 2.29 & 1.92 & -0.50 & 0.23 & 0.40 & 12.9 & 228.5 \\
1A 1742-294 & 266.52 & -29.51 & 0.69 & 2.0 & 2.47 & 5.69 $^m$ & -0.41 & 0.50 & 0.45 & 14.6 & 295.8 \\
IGR J17464-3213 & 266.56 & -32.23 & 0 & 0 & - & - & - & 1 & 2.35 & 22.3 & 1955.6 \\
GX 3+1 & 266.98 & -26.56 & 0 & 0 & - & - & - & 1 & 2.25 & 20.7 & 2237.7 \\
EXO 1745-248 & 267.02 & -24.78 & 0.33 & 0.6 & 4 & 0.04 & 0.50 & 0.39 & 0.37 & 10.4 & 197.5 \\
H 1745-203 & 267.22 & -20.36 & 0 & 0 & - & - & - & 1 & 0.79 & 6.8 & 448.7 \\
AX J1749.1-2639 & 267.30 & -26.64 & 0 & 0 & - & - & - & 1 & 1.21 & 13.2 & 1077.4 \\
SAX J1750.8-2900 & 267.60 & -29.04 & 1.84 & 2.5 & 4 & 0.18 & -0.48 & 0.21 & 0.65 & 23.3 & 419.6 \\
SWIFT J1753.5-0127 & 268.37 & -1.45 & 0.84 & 2.5 & 4 & 1.64 & -0.50 & 0.53 & 0.03 & 0.3 & 0.7 \\
 SAX J1753.5-2349 & 268.39 & -23.82 & 0 & 0 & - & - & - & 1 & 1.00 & 8.9 & 593.0 \\
4U 1755-338 & 269.67 & -33.81 & 0 & 0 & - & - & - & 1 & 1.41 & 15.7 & 1227.5 \\
GX 5-1 & 270.28 & -25.08 & 1.61 & 1.2 & 3.29 & 3.49 & -0.37 & 0.47 & 0.38 & 9.7 & 192.6 \\
GRS 1758-258 & 270.30 & -25.74 & 0 & 0 & - & - & - & 1 & 2.69 & 25.0 & 2536.6 \\
GX 9+1 & 270.38 & -20.53 & 3.85 & 1.3 & 1.51 & 11.75 $^m$ & -0.49 & 0.19 & 0.65 & 27.6 & 511.5 \\
SAX J1806.5-2215 & 271.64 & -22.25 & 0.70 & 1.6 & 2.95 & 0.35 & 0.50 & 0.45 & 0.37 & 11.3 & 201.7 \\
XTE J1807-294 & 271.75 & -29.41 & 2.81 & 3.8 & 1.87 & 0.45 $^m$ & -0.23 & 0.20 & 0.81 & 29.6 & 622.7 \\
SAX J1808.4-3658 & 272.12 & -36.98 & 0 & 0 & - & - & - & 1 & 1.89 & 17.8 & 1333.1 \\
GX 13+1 & 273.63 & -17.16 & 0.43 & 1.0 & 1.87 & 0.65 & 0.46 & 0.75 & 0.22 & 6.8 & 107.7 \\
GX 17+2 & 274.01 & -14.04 & 1.47 & 4.7 & 3.53 & 1.89 & -0.18 & 0.55 & 0.21 & 6.8 & 105.0 \\
SAX J1819.3-2525 & 274.84 & -25.43 & 0 & 0 & - & - & - & 1 & 1.38 & 15.0 & 1143.2 \\
H 1820-303 & 275.92 & -30.36 & 0 & 0 & - & - & - & 1 & 3.40 & 29.6 & 3323.0 \\
IGR J18245-2452 & 276.12 & -24.85 & 0.08 & 0.3 & 3.92 & 0.45 & -0.45 & 0.51 & 0.25 & 7.3 & 123.3 \\
Ginga 1826-238 & 277.37 & -23.80 & 2.85 & 6.2 & 3.51 & 1.32 & 0.19 & 0.21 & 0.66 & 27.1 & 472.6 \\
2S 1845-024 & 282.07 & -2.42 & 1.57 & 2.5 & 3.93 & 0.62 & -0.30 & 0.42 & 0.03 & 0.2 & 0.5 \\
IGR J18483-0311 & 282.07 & -3.18 & 0.03 & 0.7 & 4 & 0.33 & -0.48 & 0.85 & 0.04 & 0.3 & 0.8 \\
XTE J1855-026 & 283.88 & -2.61 & 3.71 & 3.8 & 4 & 0.48 & -0.20 & 0.21 & 0.05 & 0.4 & 1.0 \\
XTE J1858+034 & 284.65 & 3.44 & 0.96 & 1.5 & 4 & 0.65 & 0.28 & 0.45 & 0.03 & 0.1 & 0.3 \\
XTE J1859+083 & 284.77 & 8.25 & 0.09 & 1.0 & 4 & 0.56 & -0.43 & 0.56 & 0.04 & 0.2 & 0.6 \\
HT 1900.1-2455 & 285.04 & -24.92 & 0 & 0 & - & - & - & 1 & 2.24 & 19.1 & 2008.2 \\
XTE J1908+094 & 287.22 & 9.38 & 6.11 & 1.8 & 4 & 1.50 & -0.13 & 0.03 & 0.06 & 0.3 & 0.6 \\
4U 1907+09 & 287.41 & 9.83 & 5.54 & 2.8 & 2.26 & 1.23 & -0.40 & 0.14 & 0.05 & 0.2 & 0.6 \\
Aql X-1 & 287.82 & 0.58 & 0 & 0 & - & - & - & 1 & 0.13 & 0.3 & 3.7 \\
SS 433 & 287.96 & 4.98 & 1.15 & 8.7 & 2.71 & 0.16 & -0.19 & 0.57 & 0.05 & 0.3 & 1.1 \\
IGR J19140+0951 & 288.52 & 9.88 & 4.48 & 25.3 & 4 & 0.10 & -0.23 & 0.24 & 0.09 & 0.6 & 1.6 \\
GRS 1915+105 & 288.80 & 10.95 & 4.61 & 16.4 & 4 & 8.07 & -0.42 & 0.19 & 0.09 & 0.7 & 1.9 \\
4U 1916-053 & 289.70 & -5.24 & 0 & 0 & - & - & - & 1 & 0.26 & 0.6 & 7.8 \\
XTE J1946+274 & 296.41 & 27.37 & 3.21 & 5.8 & 2.50 & 0.00 & 0.00 & 0.09 & 0.10 & 0.5 & 1.0 \\
KS 1947+300 & 297.38 & 30.21 & 0 & 0 & - & - & - & 1 & 0.24 & 0.3 & 3.0 \\
4U 1954+31 & 298.93 & 32.10 & 0.08 & 4.0 & 4 & 0.00 & 0.25 & 0.83 & 0.08 & 0.4 & 0.9 \\
Cyg X-1 & 299.59 & 35.20 & 4.68 & 4.0 & 2.81 & 22.43 & -0.43 & 0.17 & 0.08 & 0.3 & 0.7 \\
V404 Cyg & 306.02 & 33.87 & 8.28 & 5.4 & 4 & 1.11 & -0.50 & 0.01 & 0.09 & 0.4 & 0.8 \\
V2246 Cyg & 308.06 & 37.64 & 4.24 & 1.3 & 1.09 & 0.55 & -0.04 & 0.14 & 0.16 & 0.8 & 1.7 \\
Cyg X-3 & 308.11 & 40.95 & 8.36 & 21.4 & 4 & 4.61 & 0.34 & 0.09 & 0.19 & 0.9 & 2.2 \\
SAX J2103.5+4545 & 315.90 & 45.75 & 2.15 & 6.6 & 1.97 & 0.00 & -0.50 & 0.29 & 0.15 & 0.6 & 1.4 \\
V490 Cep & 324.88 & 56.99 & 0.05 & 0.4 & 2.69 & 0.29 & 0.11 & 0.61 & 0.07 & 0.2 & 0.5 \\
Cyg X-2 & 326.17 & 38.32 & 2.44 & 17.9 & 4 & 0.48 & 0.21 & 0.42 & 0.14 & 0.7 & 2.2 \\
BD +53 2790 & 331.98 & 54.52 & 1.15 & 10.3 & 4 & 0.00 & -0.34 & 0.57 & 0.11 & 0.5 & 0.9 \\
\enddata
\tablecomments{Sources for the flare analysis with their best-fitted TS, $\hat{n}_s$, $\hat{\gamma}$, threshold $\hat{f}_{th}$, time lag $\hat{T}_{lag}$ and pre-trial $p$-values. The threshold has units $10^{-2}\rm{\, count}\,\rm{cm}^{-2}\,\rm{s}^{-1}$ and the superscript $^m$ indicates the use of MAXI light curves while others use {\em{Swift}}/BAT light curves. The time lag is represented in days. The 90\% C.L. fluence ULs are parameterized as $E^2dN_{\nu_\mu+\bar{\nu}_\mu}/dE_\nu =\mathcal{F}_{\nu_{\mu}+\bar{\nu}_{\mu}}^{90\%,\;E^{-\gamma}}\left({E_\nu}/{\rm{TeV}}\right)^{-\gamma}\,\cdot 10^{-3}\;\rm{TeV\,cm^{-2}}$, where $\gamma$ indicates the assumed spectral index of a power-law spectrum. }
\label{tab:lc_results}
\end{deluxetable*}

\begin{deluxetable}{lDDDDDDDDD}[h]
\tablecaption{Time-integrated Sources and Results}
\tablewidth{9pt}
\tablehead{
\colhead{Source} & \multicolumn2c{RA [deg]} & \multicolumn2c{$\delta$ [deg]} & \multicolumn2c{TS} & \multicolumn2c{$\hat{n}_{s}$} & \multicolumn2c{$\hat{\gamma}$} & \multicolumn2c{$p$-value} & \multicolumn2c{$\Phi_{\nu_{\mu}+\bar{\nu}_{\mu}}^{90\%,\;E^{-2.0}}$} & \multicolumn2c{$\Phi_{\nu_{\mu}+\bar{\nu}_{\mu}}^{90\%,\;E^{-2.5}}$} & \multicolumn2c{$\Phi_{\nu_{\mu}+\bar{\nu}_{\mu}}^{90\%,\;E^{-3.0}}$}
}
\decimals
\startdata
LSI +61 303 & 40.13 & 61.23 & 0 & 0 & - & 1 & 0.76 & 3.16 & 7.59 \\
LS 5039 & 276.56 & -14.85 & 0.62 & 5.78 & 3.62 & 0.382 & 1.45 & 50.43 & 834.94 \\
SS 433 & 287.96 & 4.98 & 0 & 0 & - & 1 & 0.30 & 2.85 & 9.95 \\
Cyg X-3 & 308.11 &  40.95 & 6.80 & 44.58 &  3.25 & 0.009 &  1.51 &  8.60 &  24.59 \\
\enddata
\tablecomments{Sources for the time-integrated analysis with their best-fitted TS, $\hat{n}_s$, $\hat{\gamma}$ and pre-trial $p$-values.
The 90\% C.L. flux UL are parameterized as $dN_{\nu_\mu+\bar{\nu}_\mu}/dE_\nu=\phi_{\nu_\mu+\bar{\nu_\mu}}^{90\%,\;E^{-\gamma}}\left ( {E_\nu}/{\rm{TeV}} \right )^{-\gamma}\,\cdot 10^{-12}\;\rm{TeV^{-1}cm^{-1}s^{-1}}$, where $\gamma$ indicates the assumed spectral index of a power-law spectrum. } 
\label{tab:int_results}
\end{deluxetable}

\end{document}

%% file: authors.tex
\affiliation{III. Physikalisches Institut, RWTH Aachen University, D-52056 Aachen, Germany}
\affiliation{Department of Physics, University of Adelaide, Adelaide, 5005, Australia}
\affiliation{Dept. of Physics and Astronomy, University of Alaska Anchorage, 3211 Providence Dr., Anchorage, AK 99508, USA}
\affiliation{Dept. of Physics, University of Texas at Arlington, 502 Yates St., Science Hall Rm 108, Box 19059, Arlington, TX 76019, USA}
\affiliation{CTSPS, Clark-Atlanta University, Atlanta, GA 30314, USA}
\affiliation{School of Physics and Center for Relativistic Astrophysics, Georgia Institute of Technology, Atlanta, GA 30332, USA}
\affiliation{Dept. of Physics, Southern University, Baton Rouge, LA 70813, USA}
\affiliation{Dept. of Physics, University of California, Berkeley, CA 94720, USA}
\affiliation{Lawrence Berkeley National Laboratory, Berkeley, CA 94720, USA}
\affiliation{Institut f{\"u}r Physik, Humboldt-Universit{\"a}t zu Berlin, D-12489 Berlin, Germany}
\affiliation{Fakult{\"a}t f{\"u}r Physik {\&} Astronomie, Ruhr-Universit{\"a}t Bochum, D-44780 Bochum, Germany}
\affiliation{Universit{\'e} Libre de Bruxelles, Science Faculty CP230, B-1050 Brussels, Belgium}
\affiliation{Vrije Universiteit Brussel (VUB), Dienst ELEM, B-1050 Brussels, Belgium}
\affiliation{Department of Physics and Laboratory for Particle Physics and Cosmology, Harvard University, Cambridge, MA 02138, USA}
\affiliation{Dept. of Physics, Massachusetts Institute of Technology, Cambridge, MA 02139, USA}
\affiliation{Dept. of Physics and The International Center for Hadron Astrophysics, Chiba University, Chiba 263-8522, Japan}
\affiliation{Department of Physics, Loyola University Chicago, Chicago, IL 60660, USA}
\affiliation{Dept. of Physics and Astronomy, University of Canterbury, Private Bag 4800, Christchurch, New Zealand}
\affiliation{Dept. of Physics, University of Maryland, College Park, MD 20742, USA}
\affiliation{Dept. of Astronomy, Ohio State University, Columbus, OH 43210, USA}
\affiliation{Dept. of Physics and Center for Cosmology and Astro-Particle Physics, Ohio State University, Columbus, OH 43210, USA}
\affiliation{Niels Bohr Institute, University of Copenhagen, DK-2100 Copenhagen, Denmark}
\affiliation{Dept. of Physics, TU Dortmund University, D-44221 Dortmund, Germany}
\affiliation{Dept. of Physics and Astronomy, Michigan State University, East Lansing, MI 48824, USA}
\affiliation{Dept. of Physics, University of Alberta, Edmonton, Alberta, Canada T6G 2E1}
\affiliation{Erlangen Centre for Astroparticle Physics, Friedrich-Alexander-Universit{\"a}t Erlangen-N{\"u}rnberg, D-91058 Erlangen, Germany}
\affiliation{Physik-department, Technische Universit{\"a}t M{\"u}nchen, D-85748 Garching, Germany}
\affiliation{D{\'e}partement de physique nucl{\'e}aire et corpusculaire, Universit{\'e} de Gen{\`e}ve, CH-1211 Gen{\`e}ve, Switzerland}
\affiliation{Dept. of Physics and Astronomy, University of Gent, B-9000 Gent, Belgium}
\affiliation{Dept. of Physics and Astronomy, University of California, Irvine, CA 92697, USA}
\affiliation{Karlsruhe Institute of Technology, Institute for Astroparticle Physics, D-76021 Karlsruhe, Germany }
\affiliation{Karlsruhe Institute of Technology, Institute of Experimental Particle Physics, D-76021 Karlsruhe, Germany }
\affiliation{Dept. of Physics, Engineering Physics, and Astronomy, Queen's University, Kingston, ON K7L 3N6, Canada}
\affiliation{Dept. of Physics and Astronomy, University of Kansas, Lawrence, KS 66045, USA}
\affiliation{Department of Physics and Astronomy, UCLA, Los Angeles, CA 90095, USA}
\affiliation{Centre for Cosmology, Particle Physics and Phenomenology - CP3, Universit{\'e} catholique de Louvain, Louvain-la-Neuve, Belgium}
\affiliation{Department of Physics, Mercer University, Macon, GA 31207-0001, USA}
\affiliation{Dept. of Astronomy, University of Wisconsin{\textendash}Madison, Madison, WI 53706, USA}
\affiliation{Dept. of Physics and Wisconsin IceCube Particle Astrophysics Center, University of Wisconsin{\textendash}Madison, Madison, WI 53706, USA}
\affiliation{Institute of Physics, University of Mainz, Staudinger Weg 7, D-55099 Mainz, Germany}
\affiliation{Department of Physics, Marquette University, Milwaukee, WI, 53201, USA}
\affiliation{Institut f{\"u}r Kernphysik, Westf{\"a}lische Wilhelms-Universit{\"a}t M{\"u}nster, D-48149 M{\"u}nster, Germany}
\affiliation{Bartol Research Institute and Dept. of Physics and Astronomy, University of Delaware, Newark, DE 19716, USA}
\affiliation{Dept. of Physics, Yale University, New Haven, CT 06520, USA}
\affiliation{Dept. of Physics, University of Oxford, Parks Road, Oxford OX1 3PU, UK}
\affiliation{Dept. of Physics, Drexel University, 3141 Chestnut Street, Philadelphia, PA 19104, USA}
\affiliation{Physics Department, South Dakota School of Mines and Technology, Rapid City, SD 57701, USA}
\affiliation{Dept. of Physics, University of Wisconsin, River Falls, WI 54022, USA}
\affiliation{Dept. of Physics and Astronomy, University of Rochester, Rochester, NY 14627, USA}
\affiliation{Department of Physics and Astronomy, University of Utah, Salt Lake City, UT 84112, USA}
\affiliation{Oskar Klein Centre and Dept. of Physics, Stockholm University, SE-10691 Stockholm, Sweden}
\affiliation{Dept. of Physics and Astronomy, Stony Brook University, Stony Brook, NY 11794-3800, USA}
\affiliation{Dept. of Physics, Sungkyunkwan University, Suwon 16419, Korea}
\affiliation{Institute of Basic Science, Sungkyunkwan University, Suwon 16419, Korea}
\affiliation{Institute of Physics, Academia Sinica, Taipei, 11529, Taiwan}
\affiliation{Dept. of Physics and Astronomy, University of Alabama, Tuscaloosa, AL 35487, USA}
\affiliation{Dept. of Astronomy and Astrophysics, Pennsylvania State University, University Park, PA 16802, USA}
\affiliation{Dept. of Physics, Pennsylvania State University, University Park, PA 16802, USA}
\affiliation{Dept. of Physics and Astronomy, Uppsala University, Box 516, S-75120 Uppsala, Sweden}
\affiliation{Dept. of Physics, University of Wuppertal, D-42119 Wuppertal, Germany}
\affiliation{DESY, D-15738 Zeuthen, Germany}

\author[0000-0001-6141-4205]{R. Abbasi}
\affiliation{Department of Physics, Loyola University Chicago, Chicago, IL 60660, USA}

\author[0000-0001-8952-588X]{M. Ackermann}
\affiliation{DESY, D-15738 Zeuthen, Germany}

\author{J. Adams}
\affiliation{Dept. of Physics and Astronomy, University of Canterbury, Private Bag 4800, Christchurch, New Zealand}

\author[0000-0003-2252-9514]{J. A. Aguilar}
\affiliation{Universit{\'e} Libre de Bruxelles, Science Faculty CP230, B-1050 Brussels, Belgium}

\author[0000-0003-0709-5631]{M. Ahlers}
\affiliation{Niels Bohr Institute, University of Copenhagen, DK-2100 Copenhagen, Denmark}

\author{M. Ahrens}
\affiliation{Oskar Klein Centre and Dept. of Physics, Stockholm University, SE-10691 Stockholm, Sweden}

\author[0000-0002-9534-9189]{J.M. Alameddine}
\affiliation{Dept. of Physics, TU Dortmund University, D-44221 Dortmund, Germany}

\author{A. A. Alves Jr.}
\affiliation{Karlsruhe Institute of Technology, Institute for Astroparticle Physics, D-76021 Karlsruhe, Germany }

\author{N. M. Amin}
\affiliation{Bartol Research Institute and Dept. of Physics and Astronomy, University of Delaware, Newark, DE 19716, USA}

\author{K. Andeen}
\affiliation{Department of Physics, Marquette University, Milwaukee, WI, 53201, USA}

% ...
